\documentstyle[fleqn,run2col,epsfig]{article}
\def\beq{\begin{equation}}
\def\eeq{\end{equation}}  
\def\eq#1{{Eq.~(\ref{#1})}}

\newcommand{\AmS}{{\protect\the\textfont2
  A\kern-.1667em\lower.5ex\hbox{M}\kern-.125emS}}

\title{
Cost of Survival  for  Large Rapidity Gaps
}

\author{Eugene  Levin \address{
 HEP Department, School of Physics,
 Tel Aviv University, Tel Aviv 69978, ISRAEL}
\address{
 Physics Department,
 Brookhaven National Laboratory,
 Upton, NY 11973 - 5000, USA
}%
        \thanks{Talk at ``RunII QCD and weak boson WS" March 4 - 6,
Fermilab}
        \thanks{Email: leving@post.tau.ac.il, elevin@quark.phy.bnl.gov}
 }       
\begin{document}

\begin{abstract}
In this note we report on calculations of the survival probability of the
large rapidity gap (LRG) processes and its energy behaviour.

\end{abstract}

\maketitle

\begin{flushright}
BNL-NT-99/10 \\
TAUP-2614-99\\
\end{flushright}

\section{INTRODUCTION}
In this note we consider  reaction
\beq \label{R1}
p + p  \longrightarrow
\eeq
$$
  X_1 + jet_1 (y_1, p_{1,t} \ll \mu) +  [LRG] +
jet_2 ( p_{2,t} \ll \mu) + X_2 \,\,,
$$
where LRG denotes the large rapidity gap between produced particles and
$X$ corresponds to a system of hadrons with masses much smaller than the
total energy.

The story of LRG processes started from Refs. \cite{D1,D2,D3}, where it was
noticed that these processes  give us  a  unique way to measure
high energy asymptotic  at short distances. Indeed, at first sight
 the experimental 
observable
\beq \label{D1}
 f_{gap}\; = \;  \frac{\sigma ( \;\; dijet\;\;
production \;\; with \;\;
LRG\,\,)}{\sigma_{inclusive} (\;\;  dijet \;\; production\,\,)}
\eeq
is directly related to the so called ``hard'' Pomeron exchange.
However, this is not the case and the factor ( survival probability
$\langle \mid S \mid^2 \rangle$
 appears between
the ``hard'' Pomeron exchange and the experimental observable.
\begin{figure}[h]
    \begin{center}
      \leavevmode
\epsfig{file=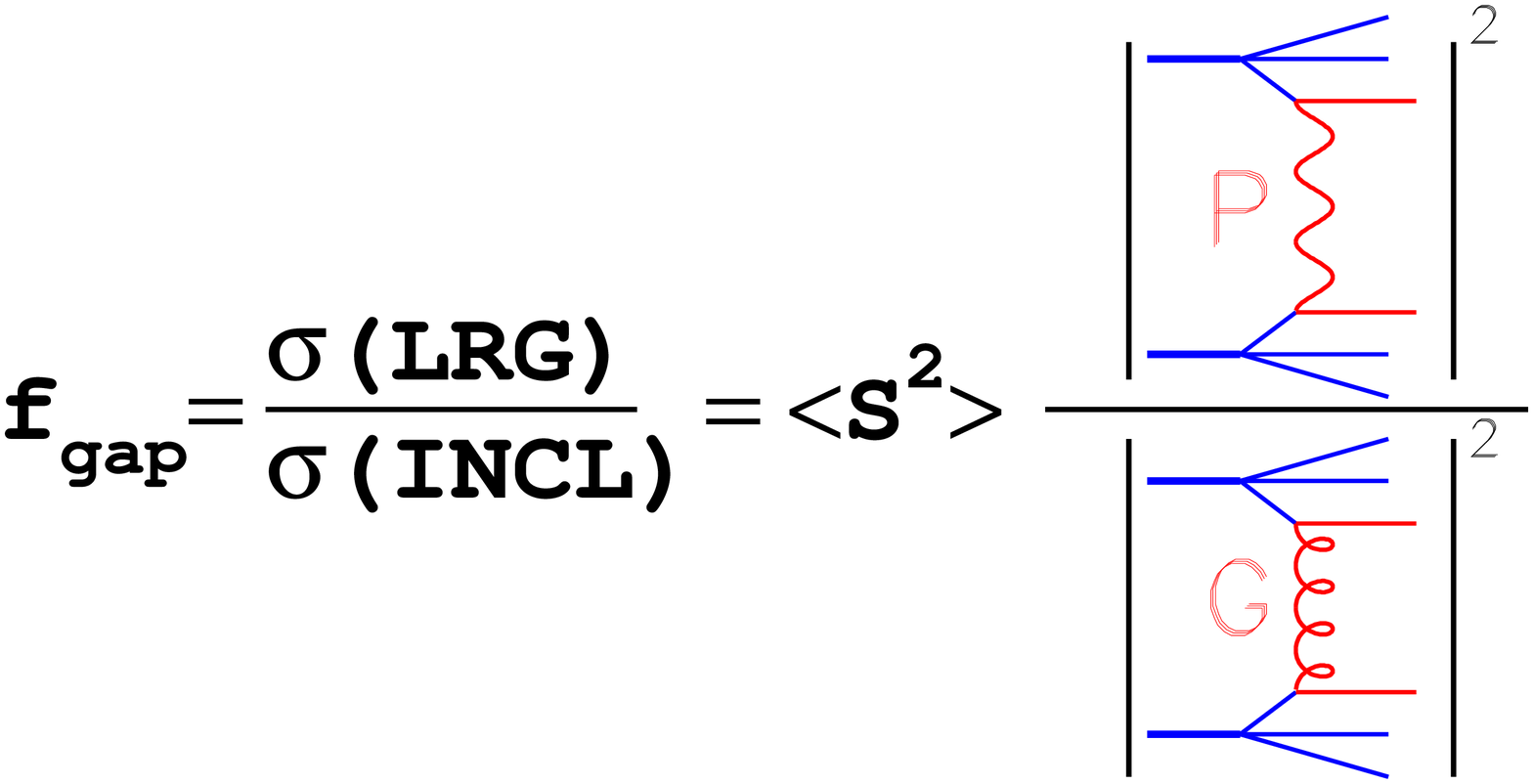,width=7cm,height=3.5cm}\\
      \label{lrg}
    \end{center}
  \end{figure}

Actually, this factor $\langle \mid S \mid^2 \rangle$ is a product of 
two survival probabilities
\beq \label{D2}
\langle \mid S \mid^2 \rangle = 
\eeq
$$
\langle\mid
S_{bremsstrahlung}(\Delta y = |y_1 -
y_2 | \mid^2 \rangle \times  \langle \mid S_{spectators}(s) \mid^2
\rangle
$$
which have different meanings.
\begin{enumerate}
\item\,\,\, $\langle \mid S_{bremsstrahlung} \mid^2 \rangle $ is
{ \it probability that the LRG will not be filled by emission of
bremsstrahlung gluons from partons, taking part in the ``hard"
interaction} ( see fig 1-a).  This factor is certainly important and
has been  calculated in pQCD in Refs. \cite{SPB1,SPB2,SPS5}. We are not
going to discuss it here;

\item\,\,\,$ \langle \mid S_{spectator}| \mid^2 \rangle$ is related to {\it
probability that every parton with $x_i\,\,>\,\,x_1$ will have no
inelastic interaction with any parton with $x\,\,<\,\,x_2$} ( see
fig. 1-b).  The situation with our knowledge of this survival
probability is the main goal of this paper.
\end{enumerate}
\begin{figure}[h]
\begin{tabular}{c }
\epsfig{file=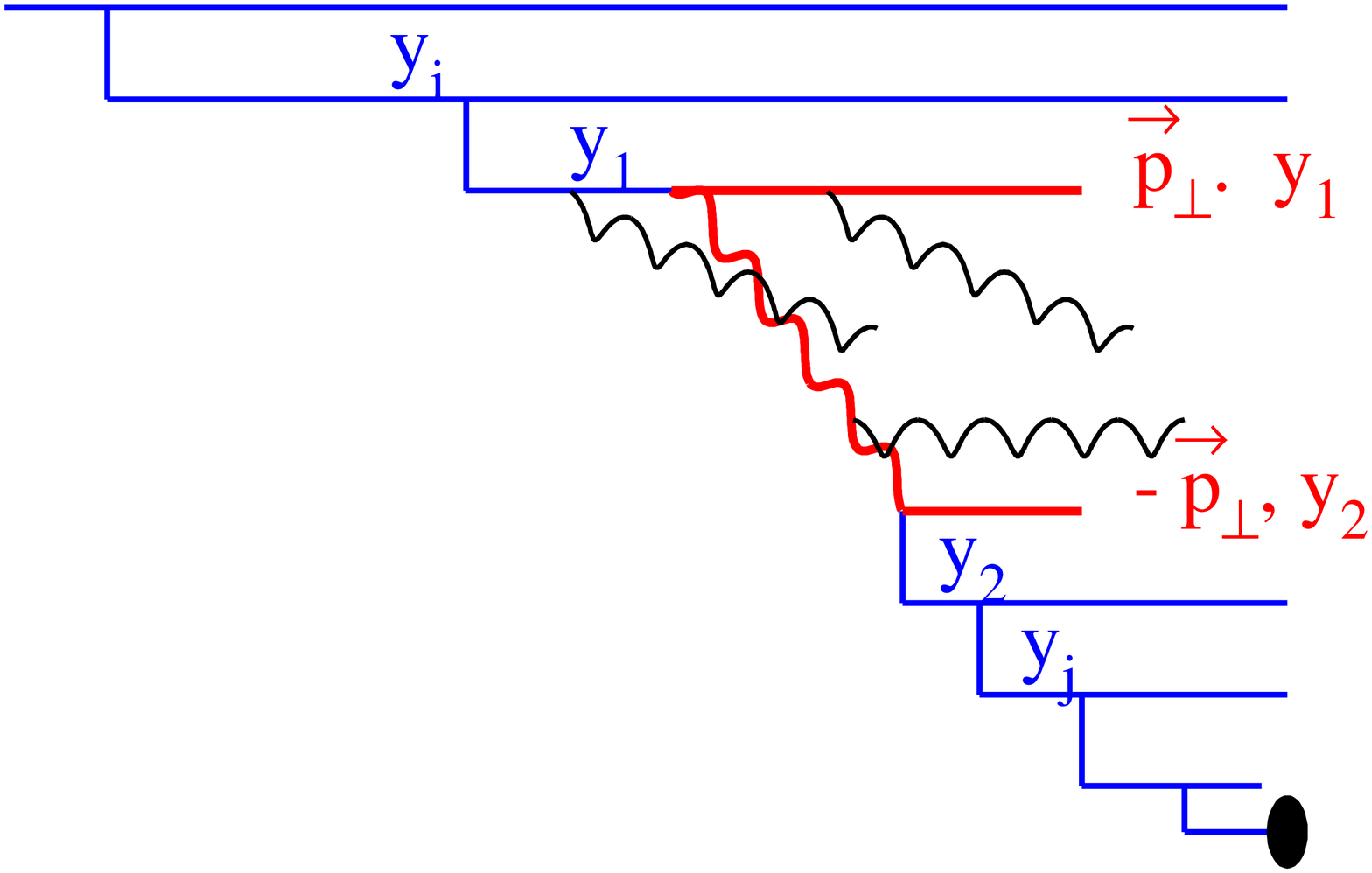,width=7cm,height=4.5cm}\\
 Fig.1-a\\
\epsfig{file=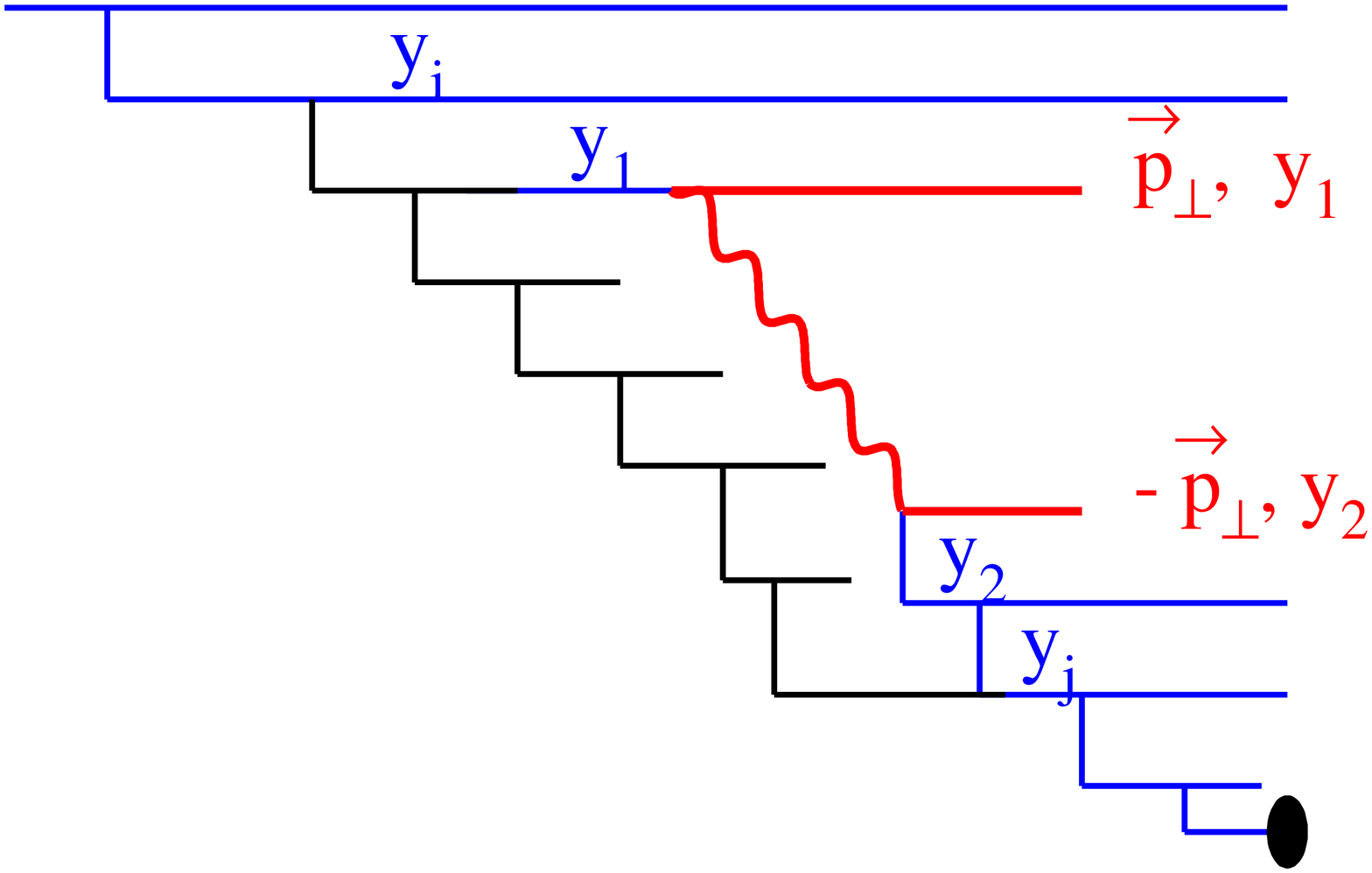,width=7cm,height=4.5cm}\\
Fig. 1-b
\end{tabular}
\caption{\it Two sources of survival probability: (a) emission of gluons from 
the partons taking parts in ``hard'' interaction and (b) emission due to
 ``soft'' interaction of spectator quarks ( partons ).}
\end{figure}

\section{Q \& A}
{\bf Q:}  Have we developed a theory for $ \langle \mid S_{spectators}(s)
\mid^2
\rangle$ ?

{\bf A:}  No, there are only models on the market (see
Refs. \cite{SPS1,SPS2,SPS3,SPS4,SPS5}).
 
{\bf Q:} Can we give a reliable estimates for the value of $\langle \mid
S_{spectator} \mid^2 \rangle$ ?

 {\bf A:}  No,  we have only rough estimates based on the
Eikonal - type models.

{\bf Q:}   Can we give a reliable estimates for the energy behaviour
of   $\langle \mid S_{spectator} \mid^2 \rangle$ ?

{\bf A:}  No, but we understood that $\langle \mid S_{spectator} \mid^2
\rangle $ could steeply decreases with energy.
 
{\bf Q:}  Why are you talking about $\langle \mid S_{spectator} \mid^2
\rangle $ if you can do nothing ?

{\bf A:}  Because dealing with models we learned what questions we should
ask
experimentalists to improve our estimate and what problems we need to
solve theoretically to provide reliable estimates.
\section{EIKONAL-TYPE MODELS}
\subsection{Eikonal model}
In eikonal model we assumed that the correct degrees of freedom at high
energies are hadrons, and, therefore, the scattering amplitude is diagonal
in the hadron basis. Practically, it means \cite{SPS1} that we assume that
the ratio $\sigma^{SD}/\sigma^{el} \,`\ll\,1$. In this model the unitarity
constraint looks simple, namely,
\beq \label{EM1}
Im a_{el}(s,b) = \mid a_{el}(s,b) \mid^{2} \,+ \, G_{in}(s,b)\,\,,
\eeq
which has solution in terms of arbitrary real function - opacity
$\Omega(s,b)$:
\begin{eqnarray} 
a_{el}\,\,&= &\,\,i\,\left[\,1 \,- \,e^{-\frac{ \Omega(s,b)}{2}}
\right]\,\,; \label{EM2} \\
G_{in}(s,b)\,\, &=&\,\, 1- e^{- \Omega(s,b)}\,\,;\label{EM3}\\
\Omega(s,b)\,\,&=& \nu(s)\,\,e^{-\frac{b^{2}}{R^{2}(s)}}\,\,;\label{EM4}
\end{eqnarray}
where \eq{EM4} is Pomeron-like parameterization that has been used for
numerical estimates. The formula for survival probability looks as
\cite{D3} \cite{SPS1}
\beq \label{EM5}
 <\mid S \mid^2> = \frac{ \int d^{2}b e^{-\frac{b^{2}}{R^{2}_H}}\,e^{-
\Omega(s,b)} }
{ \int d^{2}b  e^{-\frac{b^{2}}{R^{2}_H}}}
\eeq
where $R^2_H$ is radius for the hard processes. In Ref. \cite{SPS1}
the values of $R^2_H$ and $R^2(s)$ were discussed in details. The main
observation is that the experimental value of the ration
$\sigma^{el}/\sigma_{tot} $ depends only on the value of $\nu$. This
gives us a way to find the value of $\nu$ directly from the experimental
data. The result is plotted in Fig.2 and shows both the small value of the
survival probability and its sharp energy dependence.
\begin{figure}
\vspace{-0.4cm}
\begin{center}
\epsfig{file=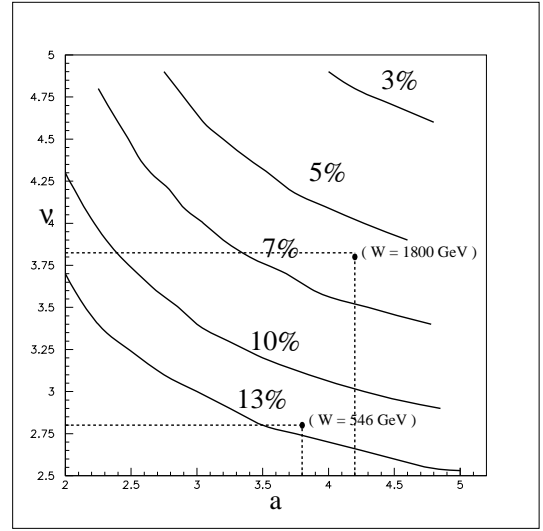,width=7cm}
\end{center} 
\caption{\it Survival probability in the eikonal model.}
\end{figure}
\subsection{Three channel model.}
The assumption that $\sigma^{SD}/\sigma^{el} \,`\ll\,1$ is in
contradiction with the experimental data, therefore, it is interesting to
generalize the eikonal model to include processes of the diffractive
dissociation. It was done in Ref.\cite{SPS2}, where the rich diffractive
final state was described by one wave function  orthogonal to the hadron
\beq \label{3CM1}
\Psi_{hadron} = \alpha\,\Psi_1 + \beta\, 
\Psi_2\,\,;\,\,\,\,\,\,\,\,\Psi_D =
-\,\beta\,\Psi_1 + \alpha\,\Psi_2 \,\,,
\eeq 
where $\alpha^2 + \beta^2 = 1$. The scattering amplitude is diagonal with
respect functions $\Psi_{1,2}$ and we used \eq{EM2}- \eq{EM4} -type
parameterization to describe it. The result of our calculation is given in
Fig.3.

\begin{figure}[t]
\begin{tabular}{c }
\epsfig{file=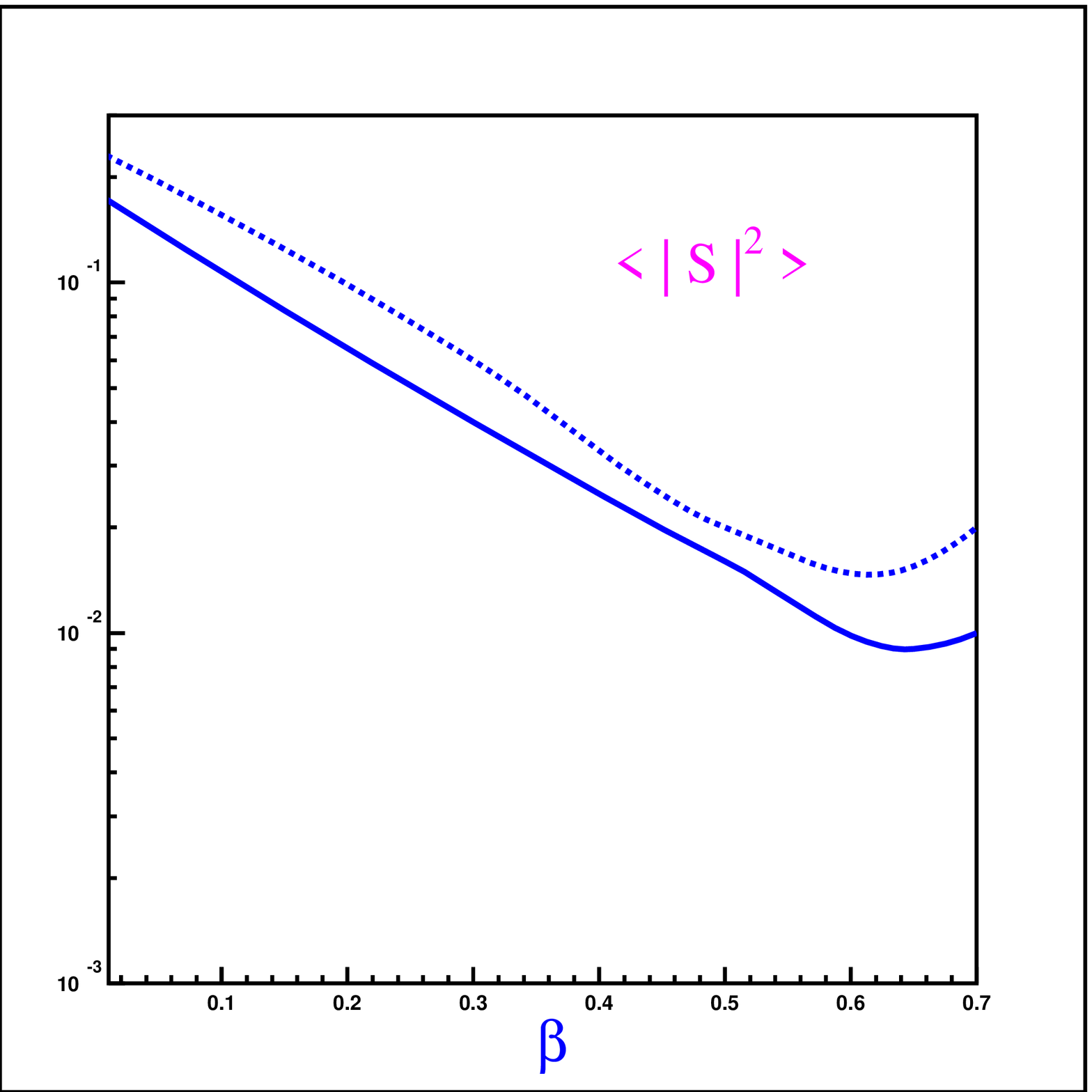,width=7cm,height=6cm}\\
 Fig.3-a\\
\epsfig{file=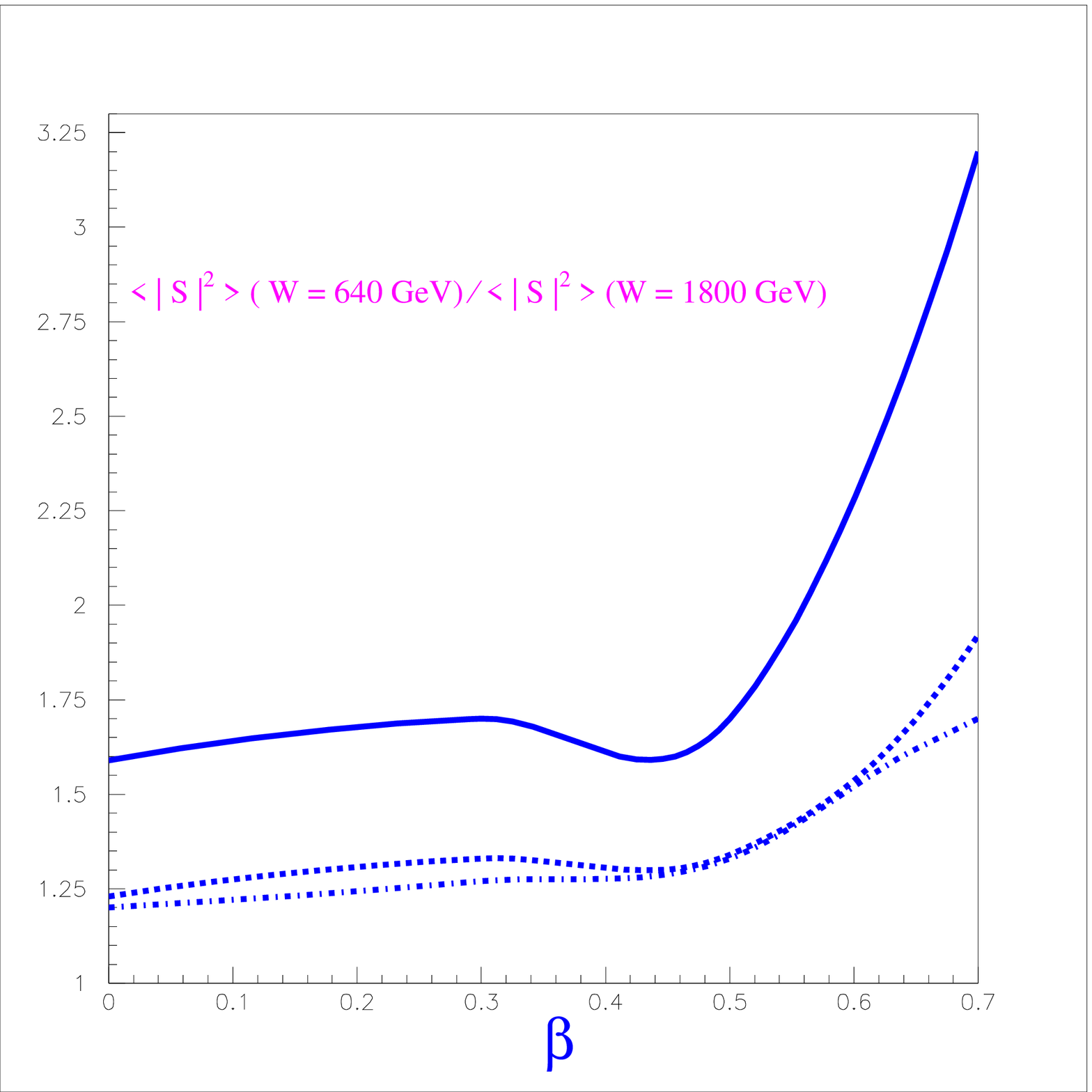,width=7cm,height=6cm}\\
Fig. 3-b\\
\epsfig{file=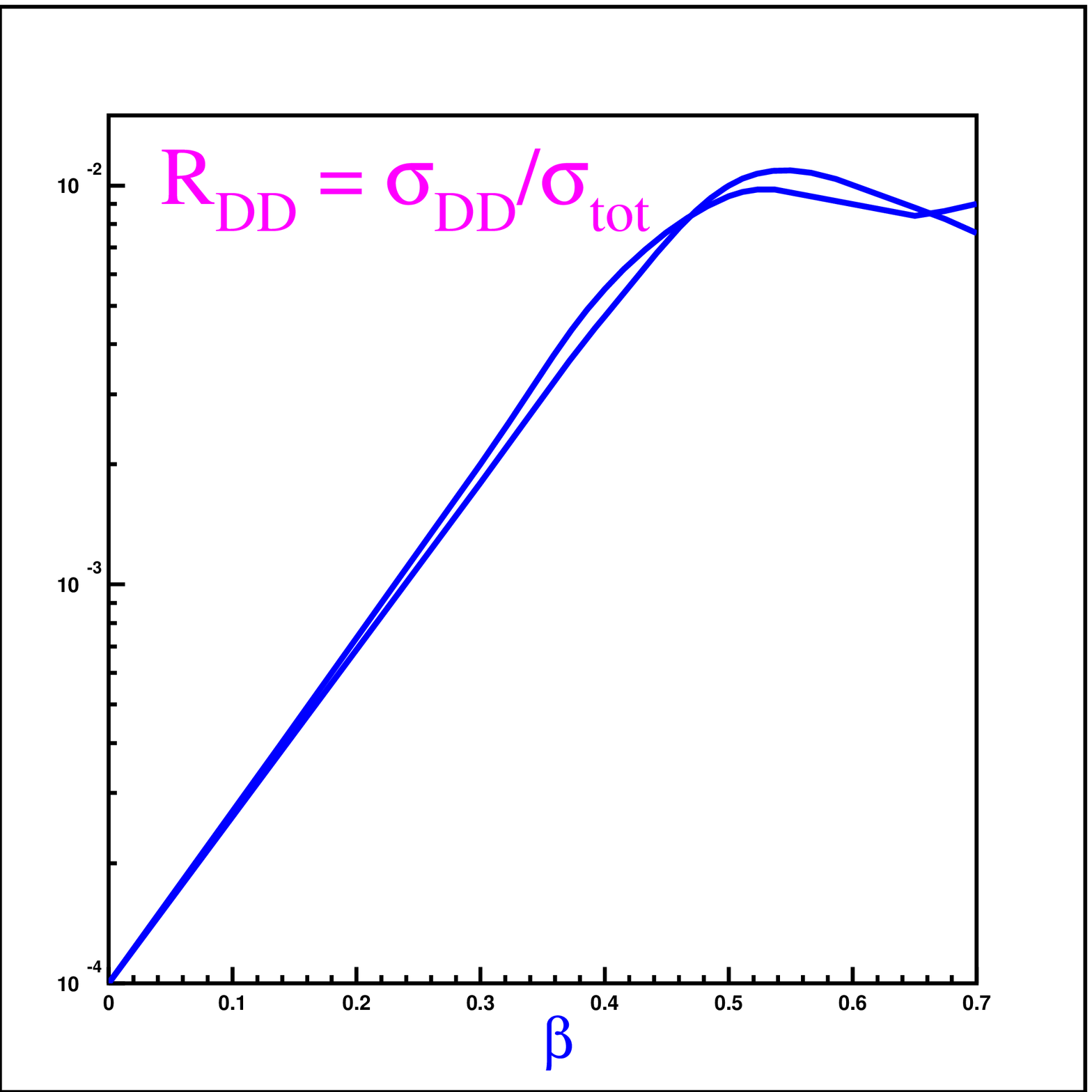,width=7cm,height=6cm}\\
Fig. 3-c
\end{tabular}
\caption{\it The value of survival probability (Fig.3-a), its energy
dependence (Fig.3-b) and prediction for the ratio of double diffraction
dissociation to the total cross section (Fig.3-c) versus $\beta$.}
 \end{figure}   
\section{CONCLUSIONS}
The experimentally observed value of the survival probability appear
naturally in these two models.

The parameters that have been used are in agreement with the more
detailed fit of the experimental data.

It turns out that the scale of $\langle \mid S_{spectator} \mid^2 \rangle$
is given by ratios
$R_{el}\,\,=\,\,\frac{\sigma_{el}}{\sigma_{tot}}$,
$R_{SD}\,\,=\,\,\frac{\sigma_{SD}}{\sigma_{tot}}$
and $R_{DD}\,\,=\,\,\frac{\sigma_{DD}}{\sigma_{tot}}$, but not 
the ratio
$R_D\,\,\,=\,\,\,\frac{\sigma_{el}\,\,+\,\,\sigma_{SD}\,\,\,+\,\,\,\sigma_{DD}}
{\sigma_{tot}}\,\,,
$ which does not show any energy dependence.

The further measurement all ratios mentioned above will specify the model
and will provide a better predictions for the survival probability. For
example, new data on  $R_{DD}$ \cite{DG} will specify the value of $\beta$
which will lead to more definite predictions for  
 $\langle \mid S_{spectator} \mid^2 \rangle$ (see Fig. 3).

It is the most dangerous that we have no theoretical approach to
calculation of the survival probability. I firmly believe that the theory
of survival for large rapidity gaps in the region of high density QCD
\cite{HDQCD} will be very instructive and will give us new ideas for the
physics of the large rapidity gap processes.

 \section*{Acknowledgements:}
I am very grateful to A. Gotsman and U. Maor for encouraging optimism and
their permanent efforts in hard business of LRG survival.
My special thanks goes to Larry McLerran and his mob at the BNL for very
creative atmosphere and hot discussion on the subject.
This  research  was supported in part by the Israel Science
Foundation, founded by the Israeli Academy of Science and Humanities,
and BSF $\#$ 9800276. This manuscript has been authorized under
Contract No. DE-AC02-98CH10886 with the U.S. Department of Energy.

\end{document}